\documentstyle[epsf]{mn}

\newif\ifAMStwofonts

\ifoldfss
  \ifCUPmtlplainloaded \else
    \NewTextAlphabet{textbfit} {cmbxti10} {}
    \NewTextAlphabet{textbfss} {cmssbx10} {}
    \NewMathAlphabet{mathbfit} {cmbxti10} {} 
    \NewMathAlphabet{mathbfss} {cmssbx10} {} 
  \fi
  \ifAMStwofonts
    \ifCUPmtlplainloaded \else
      \NewSymbolFont{upmath} {eurm10}
      \NewSymbolFont{AMSa} {msam10}
      \NewMathSymbol{\upi}     {0}{upmath}{19}
      \NewMathSymbol{\umu}     {0}{upmath}{16}
      \NewMathSymbol{\upartial}{0}{upmath}{40}
      \NewMathSymbol{\leqslant}{3}{AMSa}{36}
      \NewMathSymbol{\geqslant}{3}{AMSa}{3E}

      \let\leq=\leqslant 
       
    \fi
  \fi
\fi 

\ifnfssone
  \newmathalphabet{\mathit}
  \addtoversion{normal}{\mathit}{cmr}{m}{it}
  \addtoversion{bold}{\mathit}{cmr}{bx}{it}
  \newmathalphabet{\mathbfit} 
  \addtoversion{normal}{\mathbfit}{cmr}{bx}{it}
  \addtoversion{bold}{\mathbfit}{cmr}{bx}{it}
  \newmathalphabet{\mathbfss} 
  \addtoversion{normal}{\mathbfss}{cmss}{bx}{n}
  \addtoversion{bold}{\mathbfss}{cmss}{bx}{n}
  \ifAMStwofonts
    \ifCUPmtlplainloaded \else
      \UseAMStwoboldmath
      \makeatletter
      \new@mathgroup\upmath@group
      \define@mathgroup\mv@normal\upmath@group{eur}{m}{n}
      \define@mathgroup\mv@bold\upmath@group{eur}{b}{n}
      \edef\UPM{\hexnumber\upmath@group}
      \new@mathgroup\amsa@group
      \define@mathgroup\mv@normal\amsa@group{msa}{m}{n}
      \define@mathgroup\mv@bold\amsa@group{msa}{m}{n}
      \edef\AMSa{\hexnumber\amsa@group}
      \makeatother
      \mathchardef\upi="0\UPM19
      \mathchardef\umu="0\UPM16
      \mathchardef\upartial="0\UPM40
      \mathchardef\leqslant="3\AMSa36
      \mathchardef\geqslant="3\AMSa3E

      \let\leq=\leqslant 

    \fi
  \fi
\fi 

\ifnfsstwo
  \DeclareMathAlphabet{\mathbfit}{OT1}{cmr}{bx}{it}
  \SetMathAlphabet\mathbfit{bold}{OT1}{cmr}{bx}{it}
  \DeclareMathAlphabet{\mathbfss}{OT1}{cmss}{bx}{n}
  \SetMathAlphabet\mathbfss{bold}{OT1}{cmss}{bx}{n}
  \ifAMStwofonts
    \ifCUPmtlplainloaded \else
      \DeclareSymbolFont{UPM}{U}{eur}{m}{n}
      \SetSymbolFont{UPM}{bold}{U}{eur}{b}{n}
      \DeclareSymbolFont{AMSa}{U}{msa}{m}{n}
      \DeclareMathSymbol{\upi}{0}{UPM}{"19}
      \DeclareMathSymbol{\umu}{0}{UPM}{"16}
      \DeclareMathSymbol{\upartial}{0}{UPM}{"40}
      \DeclareMathSymbol{\leqslant}{3}{AMSa}{"36}
      \DeclareMathSymbol{\geqslant}{3}{AMSa}{"3E}

      \let\leq=\leqslant 

    \fi
  \fi
\fi 

\ifCUPmtlplainloaded \else
  \ifAMStwofonts \else 
    \def\upi{\pi}
    \def\umu{\mu}
    \def\upartial{\partial}
  \fi
\fi

\title{OGLE observations of four X-ray binary pulsars in the SMC}
\author[M.J.Coe and J. A. Orosz]
       {M.J.Coe$^{1}$ and J. A.Orosz$^{2}$ \\
$^{1}$Department of Physics and Astronomy, The University,
Southampton, SO17 1BJ, UK.\\
\\
$^{2}$Department of Astronomy and Astrophysics, 
Pennsylvania State University, 525 Davey Lab, University Park, PA
16802-6305, USA}
\date{Accepted .
      Received ;
      }

\pagerange{\pageref{firstpage}--\pageref{lastpage}}
\pubyear{1998}

\begin{document}

\maketitle

\label{firstpage}

\begin{abstract}

This paper presents analysis and interpretation of OGLE photometric
data of four X-ray binary pulsar systems in the Small Magellanic
Cloud: 1WGA J0054.9-7226, RX J0050.7-7316, RX J0049.1-7250, and 1SAX
J0103.2-7209.  In each case, the probable optical counterpart is
identified on the basis of its optical colours. In the case of RX
J0050.7-7316 the regular modulation of its optical light curve 
appears to reveal an
ellipsoidal modulation with a period of 1.416 days.  Using reasonable
masses for the neutron star and the B star, we show that the amplitude
and relative depths of the minima of the I-band light curve of RX
J0050.7-7316 can be matched with an ellipsoidal model where the B star
nearly fills its Roche lobe. For mass ratios in the range of 0.12 to
0.20, the corresponding best-fitting inclinations are about 55 degrees
or larger.  The neutron star would be eclipsed by the B star at
inclinations larger than $\approx 60^{\circ}$ for this particular mass
ratio range.  Thus RX J0050.7-7316 is a good candidate system for
further study.  In particular, we would need additional photometry in
several colours, and most importantly, radial velocity data for the B
star before we could draw more quantitative conclusions about the
component masses.

\end{abstract}


\section{Introduction}

The Magellanic Clouds present a unique opportunity to study stellar
populations in galaxies other than our own. Their structure and
chemical composition differs from that of the Galaxy, yet they are
close enough to allow study with modest sized ground based
telescopes. The study of any stellar population in an external galaxy
is of great interest because any differences with the same population
in our own Galaxy will have implications on the evolutionary
differences of the stars within the galaxies.

Be/X-ray binaries and supergiant X-ray binaries represent a subclass
of High Mass X-ray Binaries (HMXRB's). In these Be/X-ray binaries the primary
is an early type emission line star, typically 10 to 20 solar masses, 
and the secondary is a neutron star. 

The Be stars are early type luminosity class III-V stars which display
Balmer emission (by definition) and a significant excess flux at long
(IR - radio) wavelengths (dubbed as the infrared excess). These have
been successfully modelled as recombination emission (Balmer) and
free-free free-bound emission (infrared excess) from a cool dense
wind. However, observations in the ultraviolet regime indicate a
highly ionised, far less dense wind. These apparent inconsistencies
are explained in current models by assuming a non-spherically
symmetric wind structure, with a hot, low density wind emerging from
the poles of the star, and a cool, dense wind from the equatorial
regions (the circumstellar disk).

The X-ray emission is caused by accretion of circumstellar material
onto the compact companion. As a consequence, many of the Be/X-ray
binaries are known only as transient X-ray sources, with emission occurring
when accretion is enhanced during periastron passage or when the
envelope expands and reaches the neutron star. Quiescent level
emission, though low, has been detected from some of these systems.

Observations of the HMXB's in the Magellanic Clouds appear to show
marked differences in the populations. The X-ray luminosity
distribution of the Magellanic Clouds sources appears to be shifted to
higher luminosities relative to the Galactic population. There also
seems to be a higher incidence of sources suspected to contain black
holes (Clark et al. 1978; Pakull 1989; Schmidtke et al. 1994). Clark
et al. (1978) attribute the higher luminosities to the lower metal
abundance of the Magellanic Clouds, whilst Pakull (1989) refers to the
evolutionary scenarios of van den Heuvel \& Habets (1984) and de Kool
et al. (1987) which appear to favour black hole formation in low metal
abundance environments, ie. the Magellanic Clouds.

In order to study the differences between the HMXB populations of the
Magellanic Clouds and the Galaxy, it is desirable to determine the
physical parameters of as many systems as possible. We can then
investigate whether the distributions of mass, orbital period, or
spectral type are significantly different. To address these questions
we need to identify the optical counterparts to the X-ray sources
which remain unidentified, to increase the sample size to a
statistically significant number,

As part of this programme we have searched the OGLE photometric
database of 2.2 million stars (Udalski et al. 1998) to look for
counterparts to X-ray pulsars thought to be in HMXBs. Four
such systems lie within the OGLE fields and the results from binary
period searches are presented here.

\section{The OGLE data}

\begin{figure}
\begin{center}
{\epsfxsize 0.99\hsize
 \leavevmode
 \epsffile{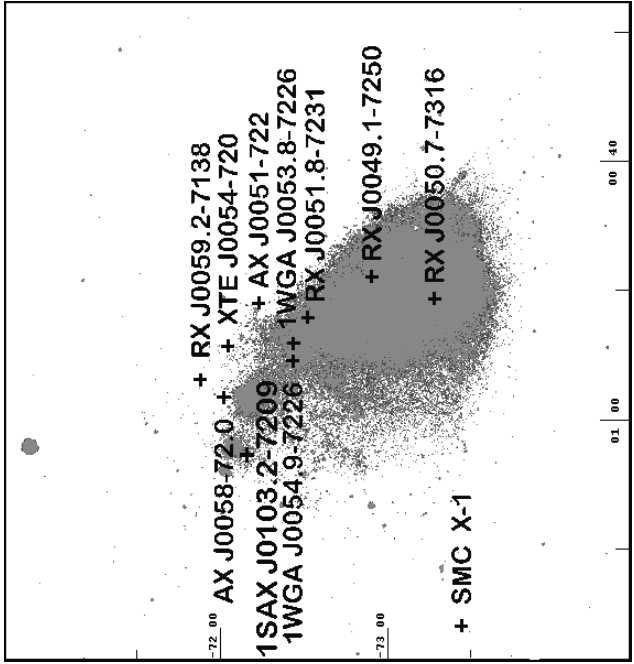}
}\end{center}
\caption{The spatial distribution of the 11 known X-ray pulsars in the
SMC.}
\label{}
\end{figure}

The Optical Gravitational Lensing Experiment (OGLE) is a long term
observational program with the main goal to search for dark, unseen
matter using the microlensing phenomenon (Udalski et al. 1992).  
The Magellanic Clouds and
the Galactic Bulge are the most natural locations to conduct such
a search due to a large number of background stars that are potential
targets for microlensing.  As a result daily photometric measurements
are made of $\sim$2.2 million stars in the SMC, and, as such, it provides
an extremely valuable resource for determining the light curves of
objects included in the survey.

Figure 1 shows the distribution of the 11 known X-ray pulsars
superimposed on an outline image of the Small Magellanic Cloud. The
OGLE scan regions are described in Udalski et al. (1998) from which it
can be seen that 7 out of the 11 pulsars lie outside the region
covered by that experiment. In fact the overall distribution of these
X-ray pulsars is far from consistent with the general visible mass
distribution of the SMC - a fact undoubtably related to the details of
star formation in this object.

In general the OGLE data cover the period June 1997 to February 1998
and primarily consist of I band observations, though some observations
were also taken in B and V. Most of the work presented in this paper
consists of identifying the possible optical counterparts to X-ray
pulsars within the OGLE database, determining their colours and
searching for regular time variability indicative of a binary period.

\subsection{1WGA J0054.9-7226}

This source was catalogued using ROSAT by Kahabka and Pietsch (1996)
and identified as an X-ray pulsar with a period of 59s by RXTE
(Marshall \& Lochner 1998). An optical study of the X-ray error boxes
clearly identified one particular object with strong H$\alpha$
emission as the counterpart to the X-ray source (Stevens, Coe \&
Buckley 1998). This object was identified within the OGLE database as
object number 70829 in Field 7 with the following magnitudes : V=15.28,
B=15.24 and I=15.13 (all magnitudes have errors of $\pm$ 0.01). Its
position, accurate to 0.1 arcsec, is RA=00 54 56.17, dec=-72 26 47.6
(2000).

\begin{figure}
\begin{center}
{\epsfxsize 0.99\hsize
 \leavevmode
 \epsffile{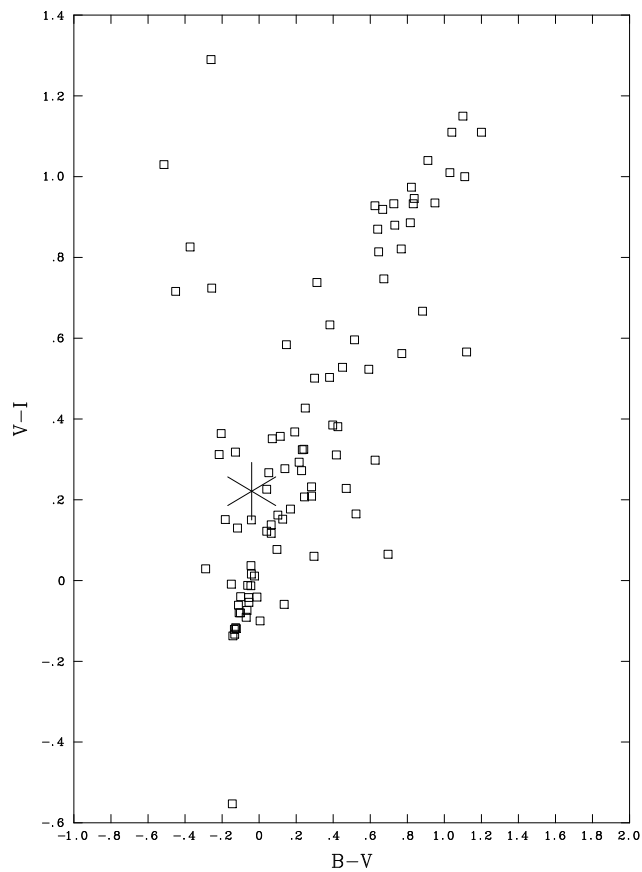}
}\end{center}
\caption{Colour-colour plots for objects in the immediate vicinity of 
1WGA J0054.9-7226. The proposed counterpart is
indicated by a star symbol.}
\label{}
\end{figure}

Its position on a colour-colour diagram is shown in Figure 2 along
with the nearest 70-80 other stars.
The position of the candidate is consistent with that expected for a
B0-B1 object. More precisely, taking V=15.28 and using the distance
modulus to the SMC of (m-M)=18.9, together with a reddening of E(B-V)
in the range 0.06-0.28 (Hill, Madore \& Freedman 1994), results in an
absolute V magnitude in the range -3.8 to -4.5. A B0V has an absolute
V magnitude of -4.0 and a B1III star has V=-4.4. In addition, it is
worth noting that the observed (B-V)= -0.04 is almost the same as that
found for another SMC X-ray counterpart RX J0117.6-7330 (Coe et al.
1998) which was identified with a star in the range B1-B2 (luminosity
class III-V). Thus one can be fairly confident in a similar spectral
class for 1WGA J0054.9-7226.

\begin{figure}
\begin{center}
{\epsfxsize 0.99\hsize
 \leavevmode
 \epsffile{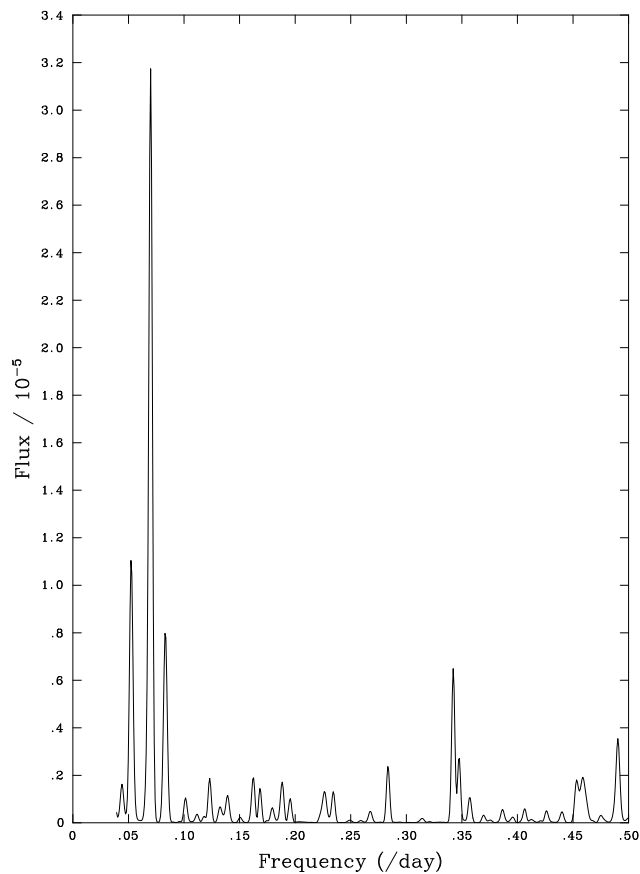}
}\end{center}
\caption{Frequency spectrum obtained by using CLEAN to analyse the
OGLE data on 1WGA J0054.9-7226. The highest peak corresponds to a
period of 14.26d.}
\label{}
\end{figure}

Figure 3 shows the power spectrum of the source obtained using CLEAN
on the 96 daily I band observation points. A strong peak occurs at a
period of 14.26d. However, when the data are folded at either that
period or twice the period, no strong light curve emerges. There is
some evidence of a small sinusoidal modulation with a semi-amplitude
of 0.008$\pm$0.001 magnitudes which presumably is the source of the
peak in the power spectrum seen in Figure 3.

\subsection{RX J0050.7-7316}

This 323s X-ray pulsar was intitially catalogued by ROSAT and
subsequently discovered to be a pulsar by Yokogawa \& Koyama (1998)
using the ASCA satellite. Subsequently Cook (1998) identified one of
the objects in the 1 armin X-ray error circle as having a period of
0.708d using data from the MACHO collaboration. Schmidtke \& Cowley
(1998) independently confirmed this identification.

In the OGLE database this source was identified as Object number
180026 in Scan Field 5. Its position is given as RA = 00 50 44.7, dec
= -73 16 05 (2000) with uncertainties of $\pm$0.1 arcsec. Its colours
are V=15.44, B=15.41 and I=15.27 with uncertainties of $\pm$0.01
magnitudes. A colour-colour plot of objects in the immediate vicinity
of the source is shown in Figure 4, from which the blue nature of the
counterpart is very clear.

\begin{figure}
\begin{center}
{\epsfxsize 0.99\hsize
 \leavevmode
 \epsffile{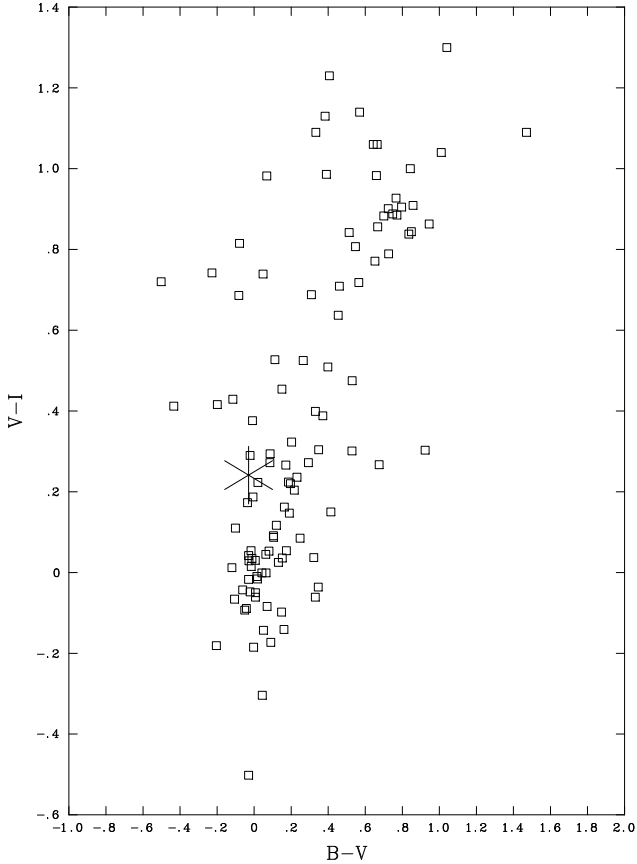}
}\end{center}
\caption{A colour-colour plot of the objects in the immediate vicinity
of the X-ray source RX J0050.7-7316. The proposed counterpart is
indicated by a star symbol.}
\label{}
\end{figure}

A CLEAN periodicity search of the 134 I band data points clearly revealed a
periodicity in the data of 0.708d - see Figure 5. This is exactly the
same as reported by Cook (1998). A full discussion of the light curve
of this source is presented in Section 3 below.

\begin{figure}
\begin{center}
{\epsfxsize 0.99\hsize
 \leavevmode
 \epsffile{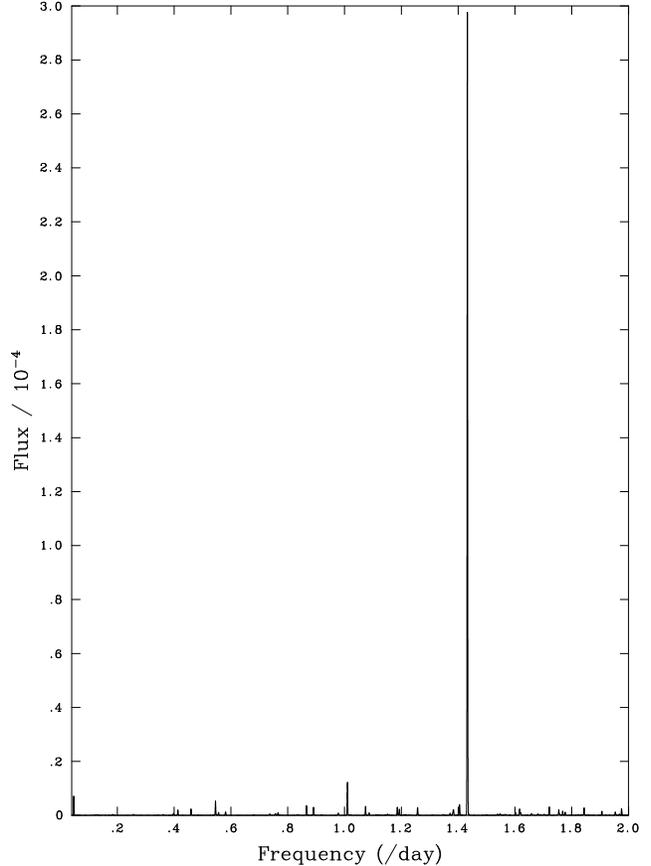}
}\end{center}
\caption{Frequency spectrum obtained by using CLEAN to analyse the
OGLE data on RX J0050.7-7316. The peak frequency corresponds to a
period of 0.708d.}
\label{}
\end{figure}

\subsection{RX J0049.1-7250}

This X-ray source was identified as an X-ray pulsar with a period of
75s by RXTE (Corbet et al. 1998) after its discovery by Kahabka
and Pietsch (1996) from ROSAT data. Stevens, Coe \& Buckley (1998) carried out a
photometric study of the objects in the X-ray error box and concluded
that Star 1 (a previously known Be star) was the most likely
counterpart. However, a second Be star (denoted Star 2 in their paper)
lies just on the edge of the ROSAT X-ray positional uncertainty and
cannot be ruled out.

Both of these objects were identified in the OGLE data base in Scan
Field 5. The data on these two objects are presented in Table 1.

\begin{table}
\caption{OGLE data on the two candidates for RX J0049.1-7250. Errors
on position are $\pm$0.1 arcsec and on magnitudes are $\pm$0.01.}
\begin{tabular}{ccc}

& Star 1& Star 2 \\
&&\\
RA (2000)& 00 49 03.3& 00 49 06.2 \\
dec (2000)& -72 50 52.0& -72 51 14.4\\
V & 16.92& 17.25\\
B & 17.01& 17.06 \\
I & 16.68& 17.18 \\
Ogle ID &65517 & 60831 \\
No. I band&146&147 \\
observations&& \\
\end{tabular}
\end{table}

\begin{figure}
\begin{center}
{\epsfxsize 0.99\hsize
 \leavevmode
 \epsffile{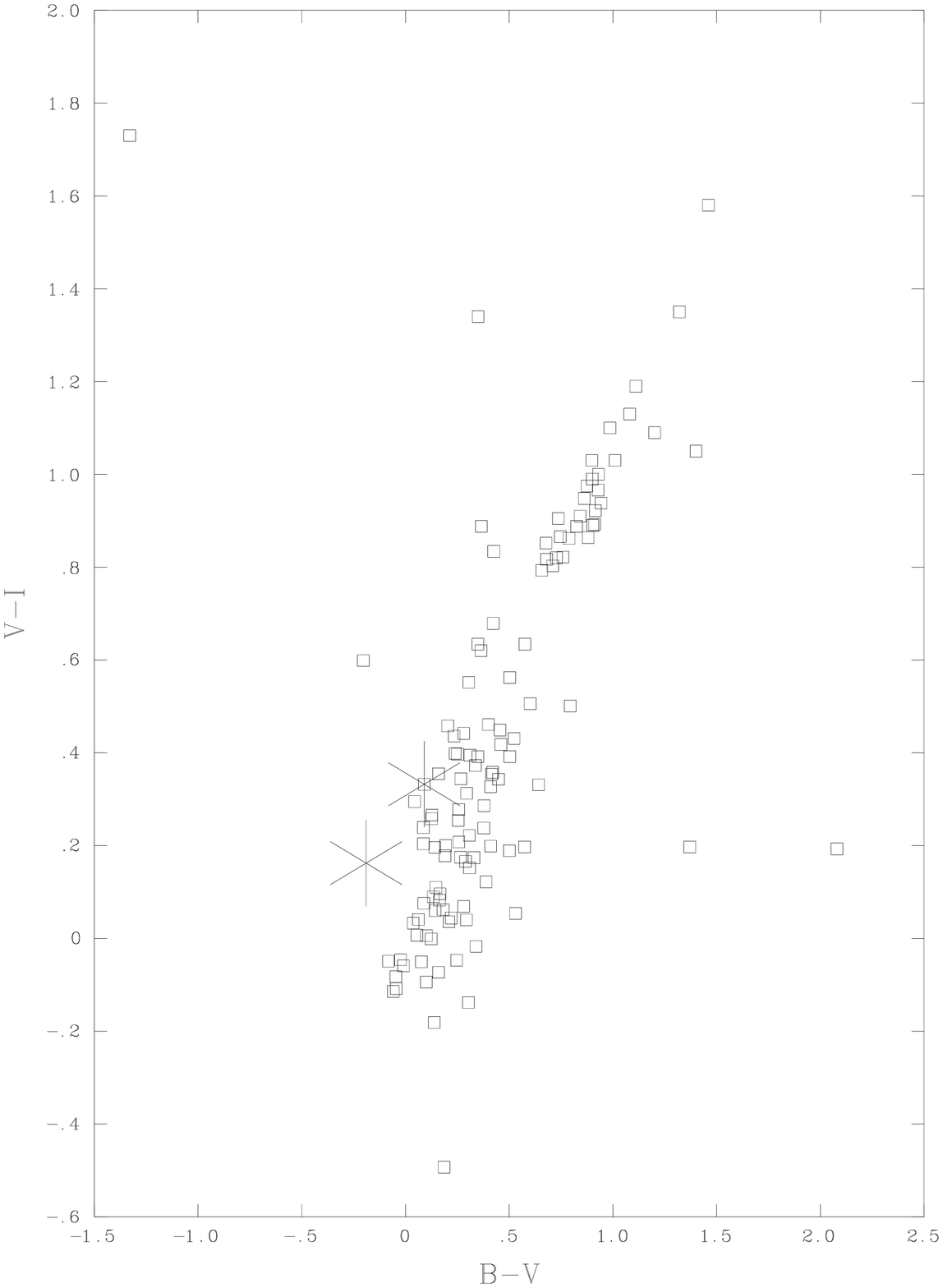}
}\end{center}
\caption{A colour-colour plot of the objects in the immediate vicinity
of the X-ray source RX J0049.1-7250.The proposed counterparts are
indicated by star symbols. Star 2 lies to the left of Star 1 in this
figure.}
\label{}
\end{figure}

A colour-colour plot for the whole of the region surrounding these two
sources is shown in Figure 6.
From this figure it can be seen that the two candidates both lie at
the blue edge of the distribution, consistent with their Be star
nature. Apart from this there appears to be nothing else remarkable
about their colours.

Both of their I band lightcurves were searched using the CLEAN
algorithm for possible periodicities similar to those found in the
previous two systems, but nothing was seen in the period range
1-50d. A modulation upper limit of $\leq$ $\pm$0.01 magnitudes may be
set based upon the signals seen from similar data runs of other
sources in this work.

\subsection{1SAX J0103.2-7209}

This object was identified in 1998 by the SAX satellite (Israel \&
Stella 1998). Its X-ray signal is modulated at a period of 345s and it
has been identified with the brightest object in the X-ray error
circle, a V=14.8 magnitude Be star. This star has previously been
proposed as the counterpart to an earlier X-ray source RX J0103-722 by
Hughes \& Smith (1994).
The proposed optical counterpart - a bright Be star - was identified
in the OGLE data base as Object number 173121 in Scan Field 9. Its
magnitudes are given as V=14.8, B-V=-0.089 and V-I=0.132 (all
magnitudes have errors of $\pm$0.01). Its position is given as RA=01
03 13.9, declination = -72 09 14.0 (2000), with a positional
uncertainty of $\pm$0.1arcsec.

\begin{figure}
\begin{center}
{\epsfxsize 0.99\hsize
 \leavevmode
 \epsffile{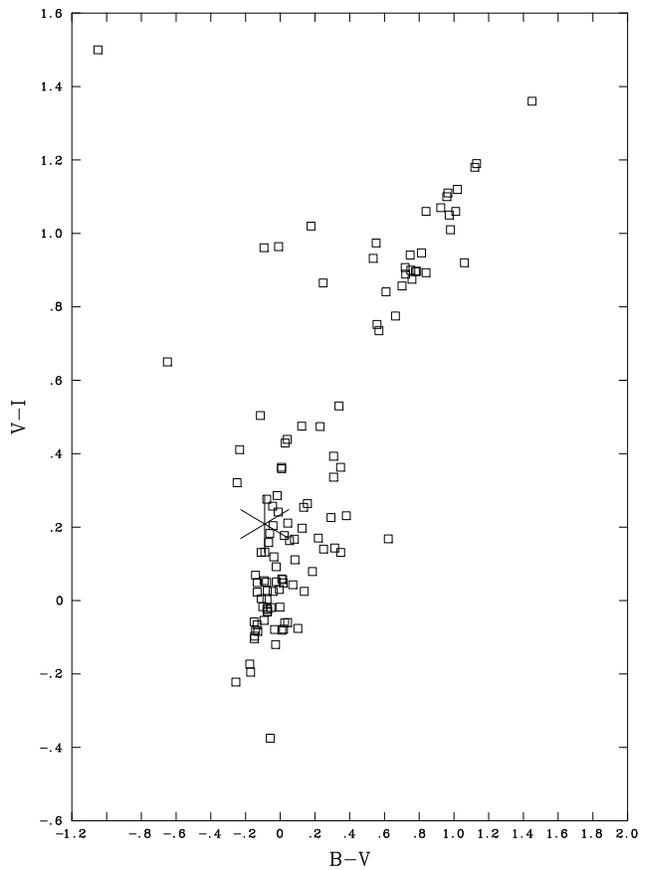}
}\end{center}
\caption{A colour-colour plot of the objects in the immediate vicinity
of the X-ray source 1SAX J0103.2-7209. The proposed counterpart is
indicated by a star symbol.}
\label{}
\end{figure}

A colour-colour plot of objects in the region around the source is
shown in Figure 7. 
The position of the proposed candidate is marked, and it is obviously
in a very similar location on this diagram to the other sources
discussed in this paper - confirming the Be star nature of this
object. There are many more objects of similar magnitude within the
SAX X-ray uncertainty circle (40 arcsec), but Hughes \& Smith (1994)
obtained a much smaller X-ray error circle ($\sim$10 arcsec) using
ROSAT which includes this Be star. It is therefore very likely that
this is the correct counterpart to the X-ray pulsar.

Timing analysis of 104 daily I band photometric measurements from OGLE
revealed no evidence for any coherent period in the range 1 to 50 days
with a modulation upper limit of $\leq$ $\pm$0.02 magnitudes.

\section{Modelling the light curve of RX J0050.7-731}

\begin{figure}
\begin{center}
{\epsfxsize 0.99\hsize
 \leavevmode
 \epsffile{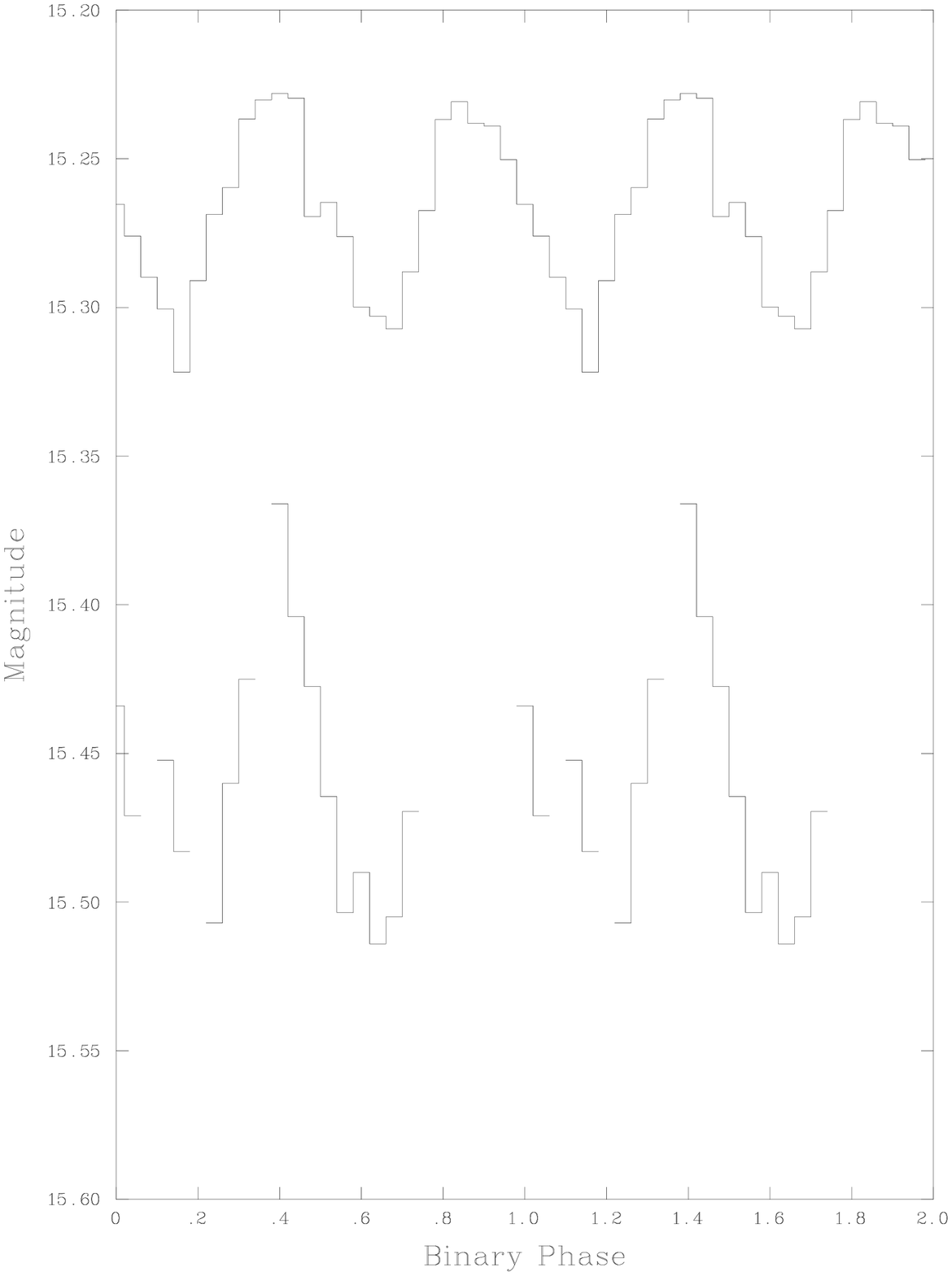}
}\end{center}
\caption{The OGLE I band data for RX J0050.7-7316 folded at the 
proposed binary period of 1.416d. The upper data are the folded I band
points and the lower data are from the V band observations.}
\label{}
\end{figure}

Using (m-M)=18.9 and E(B-V)=0.06 to 0.28, we find an absolute V
magnitude of -3.65 to -4.33 for the mass donor star of RX J0050.7-731.
This corresponds to a spectral type of B0 to B1, luminosity class
III-V.  
According to Gray
(1992), a B0V star has a mass of $13.2\,M_{\odot}$ and a radius of
$6.64\,R_{\odot}$, and a B2V star has a mass of $8.7\,M_{\odot}$ and a
radius of $4.33\,R_{\odot}$.  In order to get some idea of the scale
of this binary, we can use Eggleton's (1983) formula to compute the
relative radii of the Roche lobes
\begin{equation}
{R_{\rm L1}\over a}={0.49q^{2/3}\over 0.6q^{2/3}
+\ln(1+q^{1/3})},
\end{equation}
(where $q$ is the mass ratio), 
and Kepler's Third Law to compute the separation between the two
components (assuming a circular orbit)
\begin{equation}
a=4.2P_{\rm day}^{2/3}(M_{\rm total}/M_{\odot})^{1/3}\,R_{\odot},
\end{equation}
(where $P_{\rm day}$ is the orbital period in days).  We find $R_{\rm
L1}=3.89\,R_{\odot}$ using $P_{\rm day}=0.708$, $M_{\rm
X}=1.4\,M_{\odot}$, and $M_{\rm B}=8.7\,M_{\odot}$.  In this case
where we assumed the 0.708 day periodicity was the orbital period, the
B star clearly would overfill its Roche lobe by a large margin.  If we
assume the orbital period is 1.416 days, then we find $R_{\rm
L1}=6.17\,R_{\odot}$ using $M_{\rm X}=1.4\,M_{\odot}$, and $M_{\rm
B}=8.7\,M_{\odot}$.  We conclude the orbital period of RX J0050.7-7316
must be 1.416 days in order for the B star to fit within the Roche
lobe.  Even then, the B star still fills a large fraction of its Roche
lobe.

Folding the I and V band data for RX J0050.7-7316 at a period of 1.416
days produces the light curves shown in Figure 8.  Note that the gaps
exist in the V band data because there were much fewer observations
carried out in this band (32 compared to 134 I band measurements).
The light curves are double-wave with two maxima and two minima per
cycle.  Since the B star probably comes close to filling its Roche
lobe and is hence tidally distorted, we may presume that the cause of
the optical modulation is the well-known ellipsoidal variations (e.g.\
Avni \& Bahcall 1975; Avni 1978).  The amplitude of the ellipsoidal
light curve depends on three basic parameters: the inclination of the
orbital plane, the mass ratio, and the Roche lobe filling fraction of
the distorted star.  Thus one can in principle model ellipsoidal light
curves and obtain constraints on the system geometry.  However, the
{\em observed} light curve can be altered by the addition of extra
sources of light and by eclipses.  The amount of extra light can
depend on the orbital phase, as in light due to X-ray heating of the
secondary star, or can be independent of phase, such as light from an
uneclipsed ``steady'' accretion disk.  One must account for these
extra sources of light if one is to obtain reliable constraints from
ellipsoidal modelling.

We will focus here on only the I light curve of RX J0050.7-7316, since
the V light curve of this object is much less complete.  We used the
modified version of the Avni (1978) code described in Orosz \& Bailyn
(1997) to model the light curve.  This code uses Roche geometry to
describe the shape of the secondary star and accounts for light from a
flat, circular accretion disc and for extra light due to X-ray
heating.  The parameters for the model are the parameters which
determine the geometry of the system: the mass ratio $Q=M_x/M_B$, the
orbital inclination $i$, the Roche lobe filling factor $f$, and the
rotational velocity of the secondary star; the parameters which
determine the light from the secondary star: its polar temperature
$T_{\rm pole}$, the linearized limb darkening coefficients
$u(\lambda)$, and the gravity darkening exponent $\beta$; the
parameters which determine the contribution of light from the
accretion disc: the disc radius $r_d$, flaring angle of the rim
$\beta_{\rm rim}$, the temperature at the outer edge $T_{d}$, and the
exponent on the power-law radial distribution of temperature $\xi$,
where $T(r)=T_d(r/r_d)^{\xi}$; and parameters related to the X-ray
heating: the X-ray luminosity of the compact object $L_x$, the orbital
separation $a$ and the X-ray albedo $W$.

In this case, we can make some reasonable assumptions and greatly
reduce the number of free parameters.  We fixed the polar temperature
of the B star at 27,000~K (Gray 1992).  The B star has a radiative
envelope, so the value of the gravity darkening exponent $\beta$ was
set to 0.25 (von Zeipel 1924).  The limb darkening coefficients were
taken from Wade \& Rucinski (1985).  We can neglect the extra light
due to X-ray heating since the optical luminosity of most HMXBs 
is dominated by the mass donor star (van Paradijs \&
McClintock 1995).  We carried out some numerical experiments and found
that the light curve shapes did not depend on the values of $L_x$,
$W$, and $a$.  For definiteness we used $L_x=2\times 10^{36}$ erg
s$^{-1}$, $W$=0.5, and $a=11\,R_{\odot}$.  In that same vein, we found
that the optical flux from the accretion disk was only a small
fraction ($\la 0.5\%$) of the optical flux from the B star for a wide
range of reasonable values of $r_d$, $\beta_{\rm rim}$, $T_{d}$, and
$\xi$.  Again for definiteness we adopt $\beta_{\rm rim}=4^{\circ}$,
$T_d=6000$~K, and $\xi=-0.75$ (the value expected for a steady-state
accretion disk, Pringle 1981).  Since the accretion disk may partially
eclipse the B star, we will keep the radius of the disk $r_d$ as a
free parameter. Small changes in the disk radius can lead to
relatively large changes in the eclipse profile, whereas the eclipse
profile is much less sensitive to changes in the disk opening angle
$\beta_{\rm rim}$.  Finally, we will assume the B star is tidally
locked so that its rotational period is the same as the orbital
period.  In this case the system geometry is specified by the mass
ratio, inclination, and the Roche lobe filling factor of the B star.
Hence we have four free parameters in the model: $i$, $Q$, $f$, and
$r_d$.

We do not know the absolute orbital phase of RX J0050.7-7316, so we
simply adjusted the phase so that the deeper minimum is at phase 0.5.
We computed a grid of models in the inclination-mass ratio plane,
where the inclinations ranges from 40 to 70 degrees and the mass
ratios ranged from 0.05 to 0.35.  At each point in the plane, the
values of $i$ and $Q$ were fixed at the values corresponding to the
point in the grid and the values of $f$ and $r_d$ were adjusted to
minimize the $\chi^2$ of the fit, where we fit only the points near
the two maxima and two minima.  Figure 9 shows the contour plot of
$\chi^2$ values.  Formally, the best-fitting model has a mass ratio of
0.34 and an inclination of $50^{\circ}$.  However, the mass of the B
star probably is in the range of $\approx 7-12\,M_{\odot}$ (Gray
1992), and the mass of the neutron star probably is not too different
from $1.4\,M_{\odot}$.  Hence the mass ratio is likely to be in the
range of 0.12 to 0.20.  If we restrict our attention to fits with mass
ratios less than 0.2, we see that the corresponding best-fitting
values of the inclination are $55^{\circ}$ or higher.  We also show
the inclination at which the neutron star would be eclipsed by the B
star (thick solid line).  The model fits to the I light curve indicate
that RX J0050.7-7316 has a fairly high inclination and might show
X-ray eclipses.  It turns out that the optical light curve is altered
very little by the grazing eclipse of the star by the rim of the disk,
and of the nearly total eclipse of the disk by the star.  
Figure 10 shows a representative model light curve with
$i=60^{\circ}$, $Q=0.20$, $f=0.99$, and $r_d=0.33$.  The amplitude of
the observed I light curve is matched reasonably well, and the
relative depths of the two minima are also matched reasonably well.
However, there are relatively large deviations from the model,
especially between phases 0.4 and 0.6.

\begin{figure}
\begin{center}
{\epsfxsize 0.9\hsize
 \leavevmode
 \epsffile{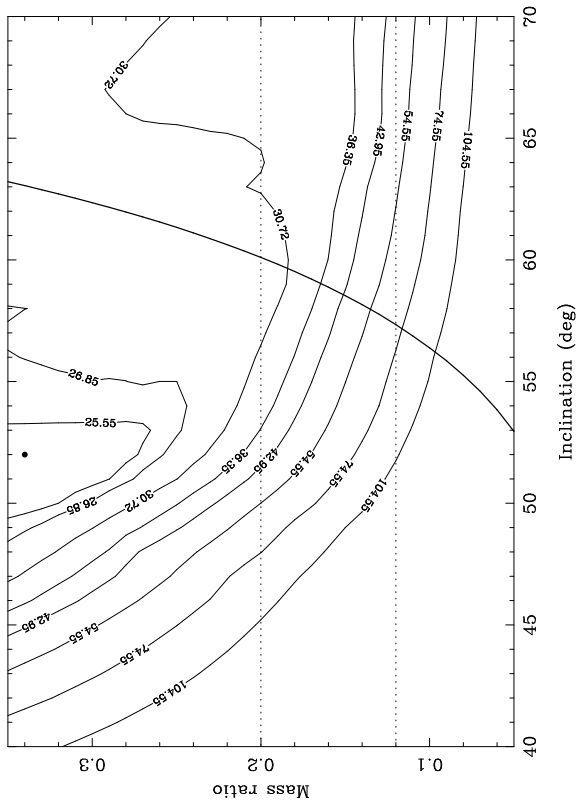}
}\end{center}
\caption{A contour plot of the $\chi^2$ values for model fitting to the folded
I light curve of RX J0050.7-7316.Eleven points near the two maxima and
two minima were included in the fitting region.  The filled circle at
$i=52^{\circ}$ and $Q=0.34$ marks the location of the lowest $\chi^2$
value ($\chi^2_{\rm min}=24.55$).
The dashed lines indicate
the likely range of mass ratios: 
$0.12\leq Q\leq 0.2$, 
which corresponds to 
$M_{\rm NS}\approx 1.4\,M_{\odot}$ 
and 
$7\leq M_{\rm B}\leq 12\,M_{\odot}$.  
The neutron star is eclipsed by the B star in models with inclinations
and mass ratios that are to the right of the thick line.}
\label{}
\end{figure}

It is difficult to draw quantitative conclusions from the modelling,
given the fact that we do not yet have dynamical data.  We have shown
that if we take the I light curve at face value, the inclination
probably is large enough to be an X-ray eclipsing system if we assume
reasonable values of the mass ratio.  There may be systematic errors
in our analysis.  For example, we assumed the OGLE light curve
represents the true ellipsoidal component from the secondary star.
Cook et al.\ (1998) report a long-term trend in the baseline
brightness in data from the MACHO collaboration.  Such a trend could
alter the amplitude of the folded light curve.  It would be worthwile
to obtain more complete photometry of this source over the course of a
several night run, rather than a few observations per night over an
entire season as in the OGLE and MACHO observations.  The light curves
obtained over a short run would be much less prone to errors
introduced by the long-term baseline brightness changes.  Furthermore,
better sampling near the minima is useful if there are grazing
eclipses since there are subtle changes in the shape of the light
curve near the minima caused by eclipses.  Finally, it would be useful
to have a radial velocity curve of the B star and a complete X-ray
light curve so that dynamical mass estimates can be derived.

\begin{figure}
\begin{center}
{\epsfxsize 0.99\hsize
 \leavevmode
 \epsffile{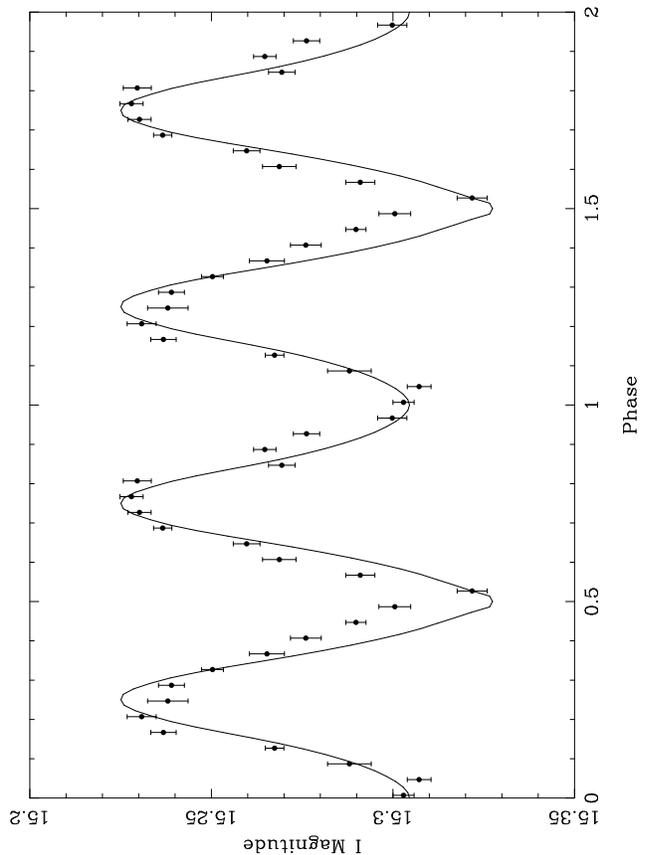}
}\end{center}
\caption{Comparison of a representative ellipsoidal model ($i=60^{\circ}$,
$Q=0.20$, $f=0.99$, and $r_d=0.33$.) and the folded I-band data for RX
J0050.7-7316.  The amplitude and relative depths of the minima are
roughly matched by the model.  However, there are relatively large
deviations from the model, especially between phases 0.4 and 0.6.
}
\label{}
\end{figure}

\section{Conclusions}

This paper presents analysis and interpretation of OGLE photometric
data of the SMC X-ray pulsars 1WGA J0054.9-7226, RX J0050.7-7316, RX
J0049.1-7250, and 1SAX J0103.2-7209.  In each case, the probable
optical counterpart is identified on the basis of its colours.  In the
case of RX J0050.7-7316, the regular modulation of its optical light
appears to reveal binary motion with a period of 1.416 days.  We show
that the amplitude and relative depths of the minima of the I-band
light curve of RX J0050.7-7316 can be matched with an ellipsoidal
model where the B star nearly fills its Roche lobe. For mass ratios in
the range of 0.12 to 0.20, the corresponding inclinations are about 55
degrees or larger. Thus the neutron star may be eclipsed by the B star
in this system.  Although the present ellipsoidal model is not
perfect, additional observations of RX J0050.7-7316 should be
obtained.  In particular, additional photometry in several colours,
and most importantly, radial velocity data for the B star will be
needed before we can draw more quantitative conclusions about the
component masses.

\section*{Acknowledgments}

We are very grateful to Andrzej Udalski and the OGLE team for their
excellent support in providing access to their data base.

\bsp

\label{lastpage}

\end{document}